\documentclass[prd,aps,superscriptaddress,nofootinbib]{revtex4}
\pdfoutput=1
\usepackage{graphicx}  
\usepackage{dcolumn}   
\usepackage{bm}        
\usepackage{amssymb}   
\usepackage[]{amsmath,amssymb}
\usepackage{graphics,epsfig}
\usepackage[latin1]{inputenc}
\usepackage[T1]{fontenc}
\usepackage{amsfonts}
\usepackage{latexsym}
\usepackage[english]{babel}
\newcommand{\be}{\begin{equation}}
\newcommand{\ee}{\end{equation}}
\newcommand{\beq}{\begin{equation}}
\newcommand{\eeq}{\end{equation}}
\newcommand{\bea}{\begin{eqnarray}}
\newcommand{\eea}{\end{eqnarray}}
\newcommand{\nn}{\nonumber}
\def\be{\begin{equation}}
\def\ee{\end{equation}}
\def\ba{\begin{eqnarray}}
\def\ea{\end{eqnarray}}

\begin{document}

\title{Spontaneous current in an holographic s+p superfluid}

\author{Ra\'{u}l E. Arias, Ignacio Salazar Landea\\
{\sl Centro At\'omico Bariloche,
8400-S.C. de Bariloche, R\'{\i}o Negro, Argentina.}}

\begin{abstract}

We study a Maxwell-Proca action in an asymptotically AdS black hole background. When moving the temperature of the black hole, we find rich phase diagrams, that depend strongly on the dimension of the operator dual to the Proca field. We present different solutions in the bulk that correspond to the holographic dual for $s$, $p$ or $s+p$-wave superfluids. In the last case we observe the onset of a spontaneously induced current.

\end{abstract}

\maketitle

\section{Introduction}

The gauge/gravity correspondence \cite{Maldacena:1997re,Witten:1998qj,Gubser:1998bc} provides a useful tool to study quantum field theories in the
strongly coupled regime. In particular, it has been used to study condensed matter systems, where a broad variety of systems are believed to be strongly coupled \cite{Hartnoll:2009sz,McGreevy:2009xe,Zaanen:2015oix}.

Among condensed matter systems, of particular interest are the high critical temperature superconductors, which have a rich phase diagram. A first attempt to approach to these interesting systems through holography was done in \cite{Hartnoll:2008vx}, where the first holographic superfluid was introduced. The model consists of a gauged complex scalar field on the background of an AdS-Schwarzschild black hole. The leading behavior of the temporal component of the gauge field close to the boundary can be associated to the chemical potential of the dual field theory. A large enough chemical potential triggers an instability for the scalar field, leading to non-trivial regular solutions. On the dual theory, this corresponds to a scalar operator acquiring a vacuum expectation value, which corresponds to a $s-$wave order parameter associated to the broken $U(1)$ global symmetry.

A first model for a holographic $ p-$wave superfluid was introduced in \cite{Gubser:2008wv} where the condensation of a vector field was modeled with an asymptotically AdS black hole having a $SU(2)$ gauge field living on it. Stringy versions for this model were studied in \cite{Ammon:2008fc,Ammon:2009fe,Kaminski:2010zu}.

An alternative model for the holographic dual to a $p-$wave superfluid was also introduced in \cite{Caiback} where, instead of a $SU(2)$ gauge field, the matter content is a complex Proca gauge field charged under $U(1)$. Again, as the temperature is lowered the normal phase becomes unstable leading to the spontaneously symmetry breaking of the $U(1)$ gauge symmetry. Since the operator that acquires a non-trivial expectation value is a vector the $SO(2)$ rotational symmetry is also broken.

In this paper we are going to study the alternative model for $p-$wave superfluidity introduced in \cite{Caiback} but considering a more general ansatz for the background fields that will allow a time-like vector condensate in the dual field theory. Since such condensate will not break the rotational symmetry, we shall call these solutions $s-$wave superfluids. In order to find such solutions, we will follow an ansatz used in the context of Proca-balls \cite{Loginov:2015rya} and Proca-stars \cite{Brito:2015pxa}. These are spherically symmetric  macroscopic condensates of a complex Proca field. We will extend the results of  \cite{Loginov:2015rya, Brito:2015pxa,Garcia:2016ldc} to gauged fields in AdS.

We will find that we can have either an $s-$wave condensate, a $p-$wave condensate or both at the same time, leading to an $s+p-$wave phase. The competition and coexistence of different order parameters is believed to play a central role in the description of the high $T_c$ superconductors \cite{Zhang} as well as other condensed matter systems \cite{Roy} and has been broadly studied in different holographic models  \cite{Zayas:2011dw,Amado:2013xya,Musso:2013rnr,Amado:2013lia,Donos:2013woa,Amoretti:2013oia,Nie:2014qma,Li:2014wca,Momeni:2013bca, Liu:2015zca,Nie:2015zia}.

This model has a new feature though: whenever two condensates coexist, a spontaneously induced current appears in the direction of the vector condensate. The apparition of spontaneous currents has been reported experimentally in high $T_c$ superconductors \cite{curr}. In this paper we will show how this spontaneous currents appear in a simple holographic model and its relation to spontaneously breaking of symmetry.

On a different context, spontaneous currents can be found on theoretical models of quark matter. Within the mean field approximation for the high density effective theory of the Kaon condensed Color Flavour Locked (CFLK)  phase of quark matter, the Nambu-Goldstone bosons coming from the spontaneous breaking of global symmetries are the relevant degrees of freedom of the theory below the chemical potential scale. This CFLK phase of quark matter has been shown to develop a spontaneous current of Nambu-Goldstone bosons due to spontaneous breaking of baryon number symmetry and hypercharge symmetry in some range of parameters \cite{Kryjevski:2008zz}.
Similar instabilities towards phases with spontaneous currents were found in models for the gluon and LOFF phases \cite{Huang:2005pv}. Even though we work with a different symmetry breaking pattern, we show that spontaneous currents may occur as well in a strongly coupled regime and beyond mean field theory.

\section{The model}

In this section we are going to set the geometric background and the matter content of our model. As was explained in the introduction we have a massive complex vector field $B_{\mu}$, charged under a $U(1)$ gauge symmetry whose gauge field is denoted by $A_\mu$. The matter action reads
\be
S=\frac{1}{2\kappa^2}\int d^4x \sqrt{-g}\left(
-\frac{1}{4}F_{\mu\nu}F^{\mu\nu}-\frac{1}{2}B_{\mu\nu}^\dag B^{\mu\nu}-m^2B_\mu^\dag B^\mu\right),\label{action}
\ee
where $F_{\mu\nu}=\partial_\mu A_\nu-\partial_\nu A_\mu$ and $B_{\mu\nu}=D_\mu B_\nu- D_\nu B_\mu$ is the field strength for the vectorial field. The covariant derivative is expressed as $D_\mu=\partial_\mu-i g A_\mu$ with $g$ the coupling between the gauge and Proca field.

We are not going to take into account the backreaction of the fields on the geometry and our choice for the metric is an AdS-Schwarzschild black hole
\be
ds^2=-f(r)dt^2+\frac{dr^2}{f(r)}+r^2(dx^2+dy^2),\label{metric}
\ee
with $f(r)=r^2\left(1-\frac{r_h^3}{r^3}\right)$ using $r_h$ to denote the position of the horizon. On the other hand, the ansatz for the matter fields reads
\be
A=\phi(r)dt+A_x(r)dx, ~~~~ B=B_t(r) dt+ i B_r(r)dr + B_x(r)dx,\label{ansatz}
\ee
for the gauge and the massive vector respectively. We will consider all fields to be real. This model was already studied in the context of holographic superfluids in \cite{Cai, Caiback} with $B_t(r)=B_r(r)=0$. We will show that with our more general ansatz new features appear.
The equations of motion read
\bea
&&\phi''(r)+\frac{2 \phi '(r)}{r}-2 \phi (r) \left(\frac{2B_x(r)^2}{r^2 f(r)}+B_r(r)^2\right)+2B_r(r)B_t'(r)+\frac{2A_x(r)B_t(r)B_x(r)}{r^2 f(r)}=0\,,\nn\\
&&A_x''(r)+\frac{A_x'(r) f'(r)}{f(r)}-2 A_x(r) \left(B_r(r)^2-\frac{B_t(r)^2}{f(r)^2}\right)-\frac{2 B_t(r)B_x(r) \phi (r)}{f(r)^2}+2B_r(r)B_x'(r)=0\,,\nn\\
&&B_t''(r)+\frac{2B_t'(r)}{r}-\frac{B_t(r)}{f(r)} \left(\frac{A_x(r)^2}{r^2}+m^2\right)-\frac{\phi (r)}{r} \left(-\frac{A_x(r) B_x(r)}{r f(r)}+r
   B_r'(r)+2 B_r(r)\right)-B_r(r) \phi '(r)=0\,,\nn\\
&&B_x''(r)
+\frac{f'(r)B_x'(r)}{f(r)}-\frac{B_x(r)}{f(r)} \left(m^2-\frac{\phi (r)^2}{f(r)}\right)-B_r(r)\left(A_x'(r)+\frac{A_x(r) f'(r)}{f(r)}\right)-A_x(r) \left(\frac{B_t(r) \phi (r)}{f(r)^2}+B_r'(r)\right)=0\,,\nn\\
&&f(r) \left(B_r(r) \left(A_x(r)^2+m^2 r^2\right)-A_x(r) B_x'(r)\right)+r^2 \phi (r) \left(B_t'(r)-B_r(r) \phi (r)\right)=0,\label{eoms}
\eea
and they have the following scaling symmetry
\be
r\rightarrow\lambda r,~~~~(B_t,B_x,\phi)\rightarrow\lambda (B_t,B_x,\phi),~~~~B_r\rightarrow \lambda^{-1}B_r\,,\label{scaling}
\ee
which allows to set $r_h=1$. We will use $g=1$ without loss of generality.

We will solve the equations of motion by shooting from the horizon out to the boundary. The gauge field and the vector must be regular at the IR. This implies the following asymptotic expansion at the horizon:
\bea
\phi(r)&\approx& \phi_h(r-1)+\frac13 \left(-2 A_h  B_{t h}  B_{x h} + \phi_h \left(-3 + B_{xh}^2 +  B_{t h}^2/m^2\right)\right)(r-1)^2+\dots \,,\nn\\
A_x(r)&\approx& A_h+ \frac1{18} \left(2 \phi_h  B_{t h} B_{x h} -A_h \left( B_{t h}^2 + B_{x h}^2 m^2\right)\right)(r-1)^2 +\dots\,  \nn\\
B_t(r)&\approx& B_{t h}(r-1) +\frac{B_{t h} \left(- \phi_h^2 + m^2 \left(-6 + A_h^2 + m^2\right)\right)}{6 m^2}(r-1)^2                           + \dots \,,\nn\\
B_x(r)&\approx& B_{x h}  +   \frac{-\phi_h A_h  B_{t h} + B_{x h} m^2 (A_h^2 + m^2)}{3 m^2}           (r-1)     +     \dots  \,,\nn\\
B_r(r)&\approx& \frac13 \left( A_h B_{x h} -\frac{\phi_h  B_{t h}}{m^2}\right) (r-1)             +  \dots \,.
\eea
We will use the undetermined horizon values $\phi_h,\, A_h,\, B_{th}
$ and $B_{xh}$ as shooting parameters to integrate our equations of motion into the desired near boundary behaviors.

The UV expansion for the matter fields reads
\bea
\phi(r)&\approx&\mu-\frac{\rho}{r},~~~~A_x(r)\approx v_x-\frac{\langle J_x \rangle}{r}\nn\\
B_t(r)&\approx& \frac{S_t}{r^{\Delta_-}}+\frac{\langle O_s \rangle}{r^{\Delta_+}},~~~~B_x(r)\approx \frac{S_x}{r^{\Delta_-}}+\frac{\langle O_p \rangle}{r^{\Delta_+}}
\eea
with $\Delta_{\pm}=\frac{1\pm\sqrt{1+4m^2}}{2}$. For a vector field the $BF$ bound establishes that $m^2>-\frac14$ \cite{McGreevy:2009xe}. According to the holographic dictionary, $\mu$ and $\rho$ will be the boundary chemical potentials and charge density and we will work at fixed chemical potential $\mu$, i.e. in the grand canonical ensemble. $\langle O_s \rangle$ and $\langle O_p \rangle$ correspond to the s and p order parameter. Since we want the $U(1)$ symmetry to be spontaneously broken, we will look for solutions with $S_t=S_x=0$. For these boundary conditions the profile for $B_r$ will be automatically regular. $J_x$ will be an spontaneously induced current that will appear for certain solutions. We will always look for solutions with $v_x=0$. Turning on a non-trivial source for $J_x$ might be interesting, and might lead into a generalization of our solutions with a possibly richer phase diagram \cite{Herzog:2008he,Basu:2008st,Arean:2010xd,Arean:2010zw,Arean:2011gz,Amado:2013aea,Wu:2014bba}. The scaling symmetry \eqref{scaling} implies that the relevant physical quantities change as
\be
T\rightarrow\lambda T,~~~~\mu\rightarrow\lambda \mu,~~~~(\rho,\langle J_x \rangle)\rightarrow\lambda^2 (\rho,\langle J_x \rangle),~~~~ (\langle O_s \rangle,\langle O_p \rangle)\rightarrow\lambda^{\Delta_+ +1}(\langle O_s \rangle,\langle O_p \rangle),
\ee
with $T=\frac{3r_h}{4\pi}$ the Gibbons-Hawking temperature of the black hole.

The free energy $\Omega$ is identified in the AdS/CFT correspondence with the on-shell action, for our ansatz it is\small
\bea
\label{feformula}
\Omega&=&\frac{1}{2}\mu\rho+\int  \frac{dr}{f(r)}\left(B_x(r)\left(f(r)^2 B_r(r) A_x'(r)+A_x(r)\left(f(r) \left(B_r(r) f(r)\right)'+2 B_t(r) \phi  \right)\right)+A_x(r)^2 \left(f(r)^2 B_r(r)^2\right.  \right.\nn \\ &&\left.\left.-B_t(r)^2\right)-r f(r)
   \left(\phi(r)\left(r B_t(r) B_r'(r)+2 B_r(r) B_t(r)+r B_r(r)^2 \phi (r)\right)+r
  B_r(r) B_t(r) \phi '(r)\right)-B_x(r)^2 \phi(r) ^2\right).
\eea
\normalsize

\section{Solutions}

In this section we present our results coming from the numerical integration of the equations of motion.
We have four kinds of solutions depending which symmetries we want to break, which will lead into four possible phases for the system.
\begin{itemize}
  \item Normal phase: here we have $B_\nu(r)=0, A_x(r)=0$ and the solution is the typical AdS-RN black hole with $\phi(r)=\mu\left(1-\frac{r_h}{r}\right)$.
  \item $s-$wave: to obtain these solutions we take non vanishing values for the fields $\phi(r)$, $B_t(r)$ and $B_r(r)$ while $B_x(r)=0$. The expectation value of a time-like operator breaks the $U(1)$ symmetry spontaneously. Since these components remain invariant under spatial rotations in the boundary we say these gravity solutions are dual to an $s-$wave superfluid.
  \item $p-$wave: the only nontrivial components of the vectors involved in this solution are $\phi(r)$ and $B_x(r)$. These solutions are dual to a $p-$wave superfluid since the order parameter is a vector that not only breaks the internal $U(1)$ symmetry but also spatial rotations.
  \item $s+p-$wave: these solutions will have a nontrivial profile for all the fields  written in \eqref{ansatz}.
\end{itemize}

Following \cite{Cai} we will focus on two particular values for the mass: $m^2=3/4, -3/16$. These values for the mass allow us to compare our results with the existing literature.  We also checked that a qualitatively similar behavior is obtained for $m^2=2,5/16, -7/64$.

\begin{figure}[h]
\begin{center}
\includegraphics[width=3.2in]{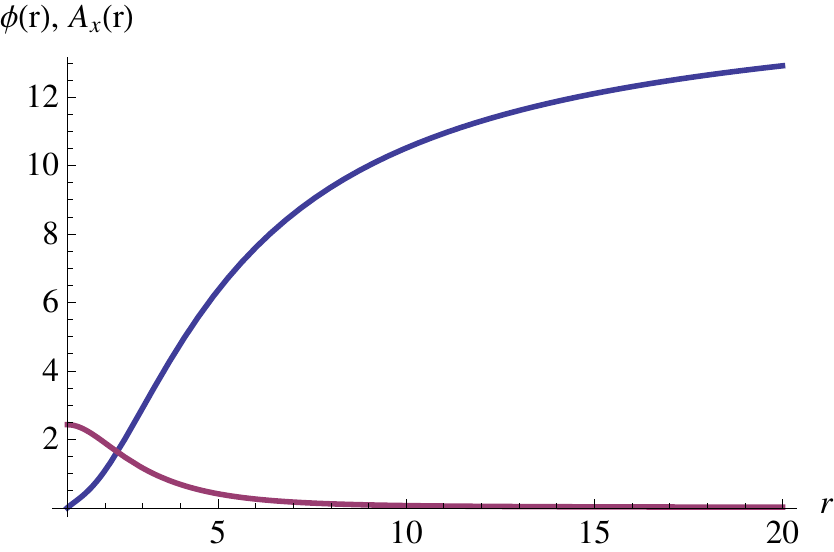}\hfill
\includegraphics[width=3.2in]{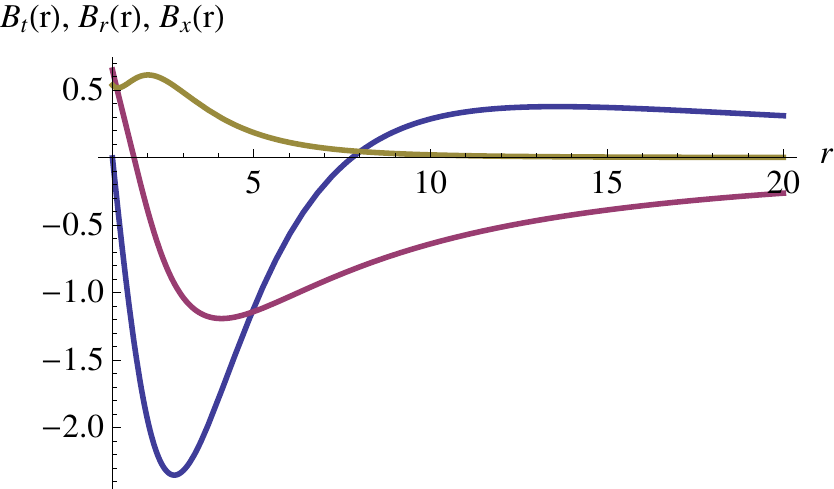}
\caption{\label{files}  Typical profile for the bulk fields in a $s+p$ superfluid solution. Left: $\phi(r)$ (blue) and $A_x(r)$ (purple). Right: $B_t(r)$ (blue), $B_r(r)$ (yellow) and $B_x(r)$ (purple). The profiles correspond to a regular solution for $m^2=3/4$ and $\mu=15.36$.  }
\end{center}
\end{figure}

Typical profiles of  the bulk fields  for the $s+p$ solution are shown in Figure \ref{files}. Notice that $B_t(r)$ has a node in the bulk (we also find solutions with more nodes, but these are all unstable). Usually one may expect that the stable solution does not have any nodes, but this is not the case for our model. This fact might be counterintuitive but it was rigourously proved for simpler configurations in \cite{Loginov:2015rya}. So far we only have numerical evidence that these solutions are indeed stable and a global minimum of the free energy.


\begin{figure}[h]
\begin{center}
\includegraphics[width=3.2in]{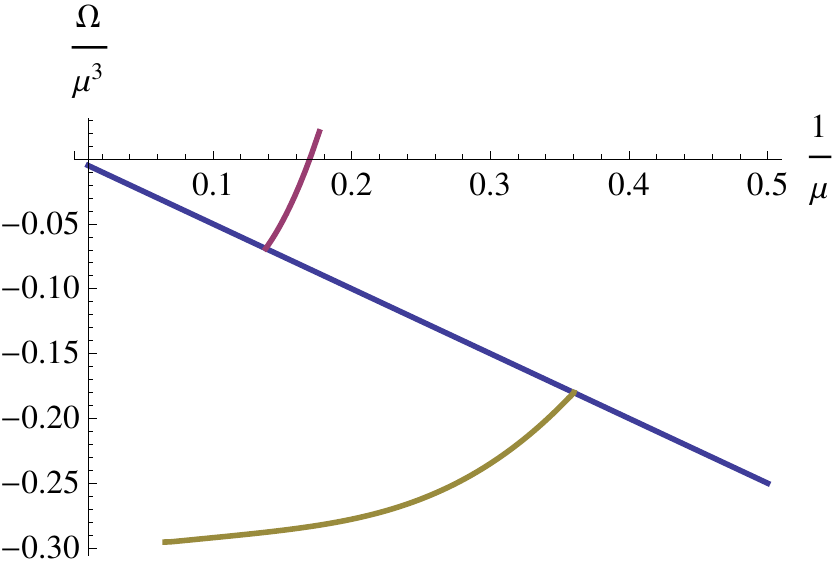}\hfill
\includegraphics[width=3.2in]{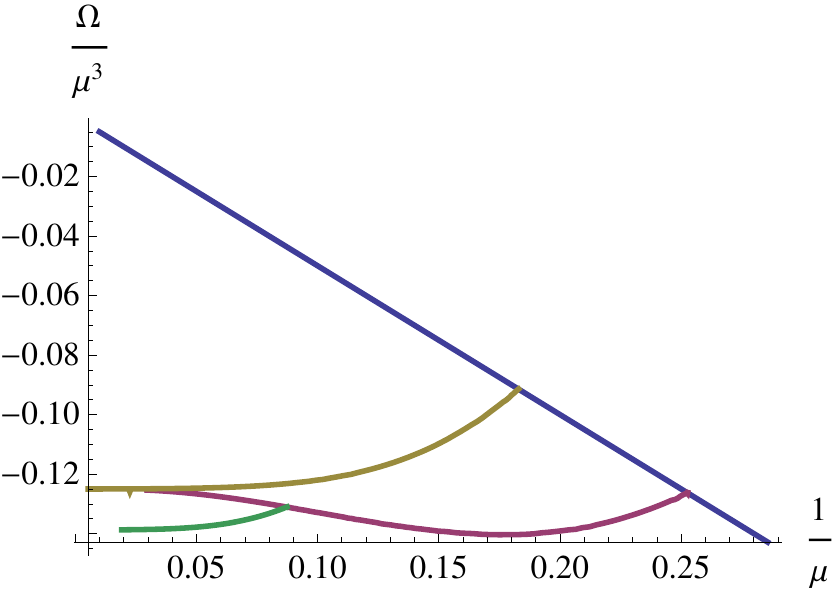}
\caption{\label{fes} Free energy as a function of the inverse chemical potential for $m^2=-3/16$ (left) and   $m^2=3/4$ (right). The different curves correspond to the normal phase (blue), p-wave (yellow), s-wave (purple) and s+p-wave (green).}
\end{center}
\end{figure}

In order to analyze the stability of the different solutions listed above we study the free energy \eqref{feformula} as a function of the chemical potential. Figure \ref{fes} shows the cases for  $m^2=-3/16$  and   $m^2=3/4$ but similar results can be obtained for other masses, finding important qualitative differences only when changing the sign of the squared mass.

For negative masses we find that the superconducting instability occurs first in the $p-$channel, when coming from the high $1/\mu$ regime. Lowering $1/\mu$ we find $s-$wave solutions but these are never energetically favoured. Then the phase diagram for this range of masses remains unchanged with respect to that presented in \cite{Cai}. In particular, for $m^2=-3/16$ we found the critical chemical potential $\mu_p\approx2.78$.

This is not the case for $m^2>0$. When lowering $1/\mu$ we find that the $s-$wave phase appears first and its energy is always lower than the $p-$wave phase. For an even lower $1/\mu$ a solution with both $s$ and $p$ order parameters appears. This solution emerges continually from the $s-$wave solutions giving a second order phase transition. In particular, for  $m^2=3/4$ we found the critical chemical potentials $\mu_s\approx3.96$ and $\mu_{s+p}\approx11.44$.

\begin{figure}[h]
\begin{center}
\includegraphics[width=3.2in]{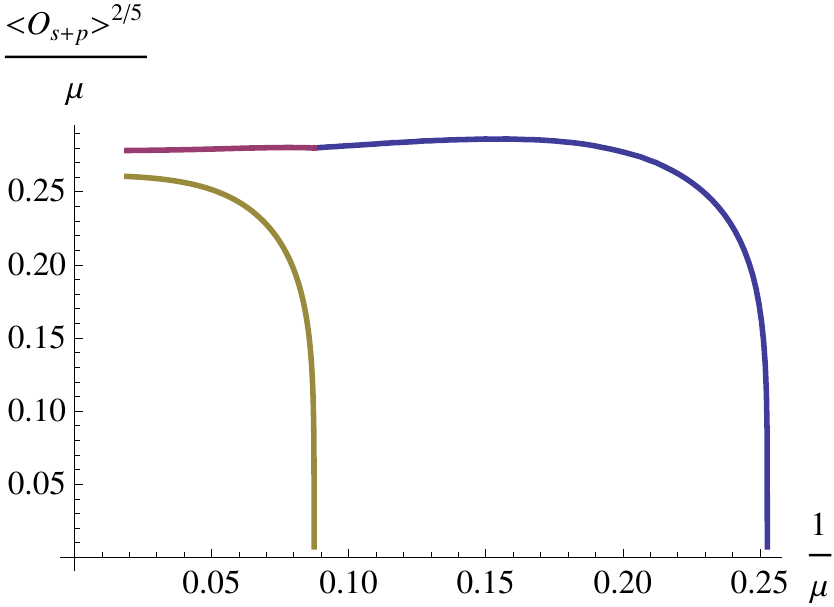}\hfill
\includegraphics[width=3.2in]{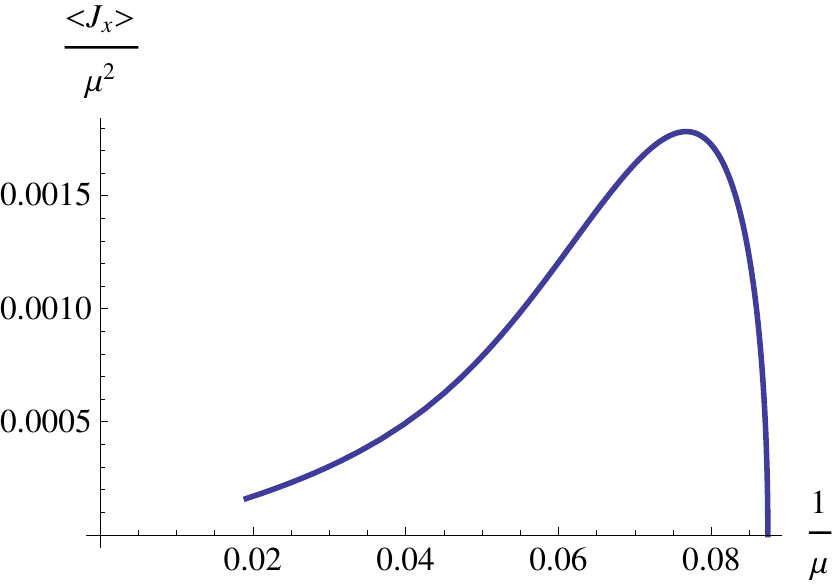}
\caption{\label{conds} Condensates (left) and spontaneously induced current (right) as a function of the inverse chemical potential for $m^2=\frac34$. The condensates correspond to the $s-$solution (blue) and the $s$ (purple) and $p$ (yellow) components of the $s+p-$condensate.}
\end{center}
\end{figure}

Finally, in Figure \ref{conds} we show some of the relevant quantities for the dual field theory in the $s+p-$wave phase, such as the condensate and the onset of an spontaneous current. Note that the spontaneous current $\langle J_x \rangle$ appears at the critical temperature where the transition from the $s-$phase to the $s+p-$phase occurs. The origin of this current can be undestood in terms of the bulk equations of motion (\ref{eoms}), since the equation of motion for the $A_x$ couples to those for the $B_{x,t}$ fields. 

\section{Conclusions}

Summarizing, we compute the thermodynamical properties associated to a Maxwell-Proca model in an asymptotically AdS black hole. We observe a rich structure of phase transitions between $s$, $p$ and $s+p$ wave superfluids. Moreover the appearance of a spontaneously generated current brings a novel interest in this kind of holographic models.

Both the apparition of spontaneous currents and the coexistence and competition of different order parameters are features of strongly correlated systems. Field theoretical studies rely on mean field approximation or perturbative methods. Here we propose an alternative approach to these new properties in the context of the AdS/CFT duality using a remarcably simple matter content. In our model the order parameters and spontaneous currents correspond to classical fields in the bulk and the couplings between them give rise to the different phases in a natural way.

As a future direction it would be interesting to realize a quasinormal modes analysis of the system. This will show the stability of our solutions and may give a useful insight about the excitations of the dual theory.
In particular, an analysis of perturbations at finite momentum will enlighten us about the existence of striped phases, for which a full non-linear analysis might be numerically more involved. The existence of striped instabilities in systems with superflows has been modeled both holographically \cite{Amado:2013aea} and field theoretically \cite{Alford:2012vn,Schmitt:2013nva,Landea:2014naa,Haber:2015exa}, even for spontaneous currents \cite{Huang:2005pv}.

\subsubsection*{Acknowledgements}
We thank to Fede Garc\'ia and Gonzalo Torroba for useful discussions. We acknowledge Daniel Are\'an for his  valuable and epic feedback.
 The authors are supported by CONICET.

\end{document}